# STOCHASTIC GENE EXPRESSION IN A LENTIVIRAL POSITIVE FEEDBACK LOOP: HIV-1 TAT FLUCTUATIONS DRIVE PHENOTYPIC DIVERSITY

Leor S. Weinberger[1]§*, John C. Burnett[3], Jared E. Toettcher[2], Adam P. Arkin[2,4]*‡, and David V. Schaffer[3]‡

Biophysics Graduate Group[1], Depts. of Bioengineering[2], The Howard Hughes Medical Institute[2] Chemical Engineering[3], and the Helen Wills Neuroscience Institute[3],
University of California, Berkeley, CA 94720
and
Physical Biosciences Division[4],
Lawrence Berkeley National Laboratory, Berkeley, CA 94720

* corresponding authors: leor@princeton.edu, Tel: (609) 258-6785, Fax: (609) 258-1704
aparkin@lbl.gov, Tel: (510) 495-2116, Fax (510) 486-6129

‡ these authors contributed equally to this work

Running Title: Stochastics in HIV-1 transactivation

Manuscript Information
50 text pages (abstract, text body, and references)
6 Figures, 1 Table, 13 Equations
Character Count (text, figure legends, methods, references): 54,548
Abstract: **143 Words**

Abbreviations:
LTR. Long Terminal Repeat; IRES, Internal Ribosomal Entry Site; LGIT, LTR-GFP-IRES-Tat (an HIV-1 derived lentiviral vector); LG, LTR-GFP (an HIV-1 derived lentiviral vector); ODE, Ordinary Differential Equation; HERV, Human Endogenous Retrovirus; RNAPII, RNA Polymerase II; MOI, Multiplicity of Infection; GFP, Green Fluorescent Protein; TNFα, Tumor Necrosis Factor α; PMA, Phorbol Myristate Acetate; TSA, Trichostatin A; SINE, Short Interspersed Nuclear Element; LINE, Long Interspersed Nuclear Element; PheB, Phenotypic Bifurcation; PTEFb, Positive Transcriptional Elongation Factor b; Cdk9, Cyclin dependent kinase 9; RFU, Relative Fluorescence Units.

SUPPLEMENTARY INFORMATION ATTACHED.

**§ Current address: Dept. of Molecular Biology, Princeton University, Princeton, NJ 08544-1014**

## SUMMARY

HIV-1 Tat transactivation is vital for completion of the viral lifecycle and has been implicated in determining proviral latency. We present an extensive experimental/computational study of an HIV-1 model vector (LTR-GFP-IRES-TAT) and show that stochastic fluctuations in Tat influence the viral latency decision. Low GFP/Tat expression was found to generate bifurcating phenotypes with clonal populations derived from single proviral integrations simultaneously exhibiting very high *and* near zero GFP expression. Although phenotypic bifurcation (PheB) was correlated with distinct genomic integration patterns, neither these patterns nor other extrinsic cellular factors (cell cycle/size, aneuploidy, chromatin silencing, etc.) explained PheB. Stochastic computational modeling successfully accounted for PheB and correctly predicted the dynamics of a Tat mutant that were subsequently confirmed by experiment. Thus, Tat stochastics appear sufficient to generate PheB (and potentially proviral latency), illustrating the importance of stochastic fluctuations in gene expression in a mammalian system.

## INTRODUCTION

Historically, genetics was considered a largely deterministic process where the presence or absence of a gene conferred alternate phenotypes. Non-deterministic genetics (i.e. variegated or mottled phenotypes) were known to exist but were limited to examples of epigenetic heterochromatin spreading, such as position effect variegation (PEV) in Drosophila (Reuter and Spierer, 1992). Recently, however, common gene regulatory motifs, such as positive and negative feedback loops, have been shown to produce variegated phenotypes based solely on stochastic fluctuations, or noise, in their molecular constituents. When gene expression is governed by low transcriptional rates and low concentrations of transcriptional complexes, small thermal fluctuations become highly significant and can drive phenotypic variability.

The effect of endogenous noise in cellular processes was predicted as early as the 1970s (Spudich and Koshland, 1976), but the physical basis of noise sources in prokaryotic gene expression was only recently theoretically analyzed (McAdams and Arkin, 1997) and directly measured (Becskei and Serrano, 2000; Elowitz et al., 2002; Isaacs et al., 2003). Studies in *S. cerevisiae* subsequently demonstrated that noise in eukaryotic transcription and translation alone could lead to population variability (Becskei et al., 2001; Blake et al., 2003; Raser and O'Shea, 2004).

Studies of a bacterial virus, λ phage, have demonstrated that the phage's lysis/lysogeny lifecycle decision is a paradigmatic example of how stochastic thermal fluctuations in chemical reaction rates can induce variability and influence a life-cycle decision. Stochastic expression from phage λ's divergent $p_R$ and $p_{RM}$ promoters is critical in controlling the λ switch and can explain the phage's non-determinic lysis/lysogeny choice (Arkin et al., 1998). Subsequent experimental measurements of expression from the λ promoter, and its operation in synthetic circuits, confirmed its fundamentally stochastic nature (Elowitz and Leibler, 2000). There is also evidence that mammalian DNA viruses, such as Simian Virus 40 and cytomegalovirus, control lysogeny (or latency) lifecycle decisions without relying on PEV-like mechanisms (Hanahan et al., 1980; Reddehase et al., 2002). Here, we explore whether latency in the retrovirus HIV-1 can be explained by stochastic gene expression.

The Human Immunodeficiency Virus type 1 (HIV-1) establishes a long-lived latent reservoir *in vivo* by persisting as a stable integrated provirus in memory $CD4^+$ T lymphocytes (Pierson et al., 2000). These latently-infected resting cells constitute a small population, are extremely long-lived, are not believed to produce appreciable virus (Finzi et al., 1999), and are considered the most significant obstacle thwarting HIV-1 eradication from a patient. While molecular determinants of HIV-1 latency have been explored (Brooks et al., 2001; Jordan et al., 2003; Kutsch et al., 2002), there has been no conclusive identification of host cellular factors that direct a T lymphocyte to become latent. The HIV-1 Tat protein, however, appears to be vital in viral progression to latency (Jordan et al., 2001; Lin et al., 2003).

After integration into a semi-random position in the human genome (Schroder et al., 2002), HIV-1 is thought to establish a low basal expression rate that depends upon the integration site (Jordan et al., 2001). Tat, Rev, and Nef proteins are then produced initially at a low basal rate from multiply spliced short transcripts. In the cytoplasm, Tat, CDK9 and CyclinT1 bind to form the positive transcriptional elongation factor b (pTEFb). pTEFb is subsequently imported into the nucleus where it is acetylated (Kaehlcke et al., 2003) and amplifies HIV-1 transcriptional elongation by releasing RNA Polymerase II (RNAPII) from its stalled position on the HIV-1 Long Terminal Repeat (LTR). Tat transactivation thus amplifies viral transcriptional elongation ~100-fold and thereby increases mRNA products from the viral genome to establish a very strong positive transcriptional feedback loop (Karn, 2000). However, integrations in some regions such as heterochromatic alphoid repeats allow virtually zero basal expression (Jordan et al., 2003). Therefore, 2 possible transcriptional modes exist: low/basal and high/transactivated.

In general, low molecular concentrations and slow reactions, which exist initially after viral infection, can exhibit large thermal fluctuations in reaction rates and molecular concentrations. Furthermore, positive feedback (e.g. Tat transactivation) can amplify these fluctuations to drive phenotypic variability. To test this hypothesis we built a lentiviral model system containing the Tat feedback loop, LTR-GFP-IRES-Tat (LGIT). Jurkat cells infected with LGIT exhibited stochastic switching between transactivated (high GFP expression) and untransactivated (low GFP expression) states, and isolated clonal populations generated a variegated expression phenotype, or *phenotypic bifurcation* (PheB). By contrast, a control virus lacking the Tat feedback loop exhibited a low, stable level of expression. *In silico* modeling indicated that stochastic fluctuations in the Tat transactivation circuit were sufficient to account for PheB, and predicted the expression dynamics of a mutated Tat circuit that was subsequently experimentally tested. Simulation and experiment therefore suggest that extended stochastic delays in viral transactivation are possible, and such a delay could generate proviral latency.

## RESULTS

### Computational design of the LTR-GFP-IRES-Tat (LGIT) lentiviral vector

Previous work suggested that an HIV-1 provirus adopts one of two mutually-exclusive expression modes (Jordan et al., 2003), a high Tat concentration transactivated mode or a low/zero Tat concentration basal mode. However, the provirus may exhibit interesting expression dynamics between these two modes. Specifically, after viral integration into regions that permit only a very low/slow basal rate, random fluctuations could be amplified by Tat positive feedback to generate stochastic pulses of Tat expression, thereby yielding a variegated expression profile. To test this hypothesis (Fig. 1a), a simplified stochastic simulation based on *in vitro* Tat transactivation dynamics was used to explore the low Tat concentration regime (Supp. Info) (Gillespie, 1977), and it revealed a probabilistic, highly unstable region between the Tat transactivated and untransactivated modes (Fig. 1b). Intermediate Tat concentrations could either be stochastically driven by positive feedback to turn the viral "circuit" on to a high expression state, or decay to turn the circuit off.

To experimentally test this preliminary prediction, an HIV-1 based lentiviral vector LTR-GFP-IRES-Tat (LGIT) (Fig. 1c) was constructed. GFP was used to quantify the activity of the HIV-1 long terminal repeat (LTR) promoter, and Tat was positioned 3' of the internal ribosomal entry sequence (IRES), somewhat analogous to its position in HIV-1 downstream of multiple splice acceptor sites. The IRES, known to reduce expression of the 2$^{nd}$ cistron at least 10-fold relative to the first cistron (Mizuguchi et al., 2000), was used to effectively enrich for low Tat concentrations, thus emulating HIV-1 integration into regions that generate low basal expression. To isolate the effect of the Tat transactivation loop, no other HIV-1 genes were included in the vector. A vector lacking Tat positive feedback (LTR-GFP, or LG) was constructed as a control.

Low Tat/GFP concentrations are unstable in LGIT infections

Jurkat cells were infected with the LG and LGIT lentiviral vectors at a low multiplicity of infection (MOI<0.1). LG-infected Jurkats expressed low GFP levels that appeared stable over many weeks (Fig. 2a). By contrast, LGIT-infected cells exhibited a population of highly fluorescent cells at levels >50-fold above background, referred to as *Bright* cells (>25 Relative

Fluorescence Units, RFU, Fig. 2c). Although >90% of cells were uninfected, most LGIT-infected cells appeared to express GFP at high levels since subsequent stimulation with tumor necrosis factor α (TNFα) produced only a minimal increase in the GFP+ population (Supp. Info.). Importantly, a small fraction of LGIT infected cells (<1%) exhibited an intermediate level of fluorescence between *Bright* and *Off*, which we refer to as *Dim* and *Mid* fluorescence (0.5 - 4 RFU and 4 - 20 RFU, respectively). Furthermore, the fraction of cells in the LGIT *Dim* region (Fig. 2c), also the low Tat concentration region (see Fig. 3h), appeared to fluctuate significantly over a period of days (Supp. Info.), consistent with the concept that intermediate expression levels may be unstable (Fig. 1b). This fluctuation or instability in the LGIT *Dim* region led us to examine the stability of all LGIT and LG fluorescence regions. Fluorescence activated cell sorting (FACS) was utilized to isolate ~10,000 cells from different regions of GFP fluorescence: *Off*, *Dim*, *Mid*, and *Bright* (Fig. 2a, c). Post-FACS analysis verified that sorting fidelity was >98% (data not shown).

The LG *Dim* and *Mid* sorted subpopulations remained completely stable for many months, though the *Mid* cells exhibited a small initial relaxation to the *Dim* region with no additional relaxation over time (Fig. 2b). Notably, no fraction of the *Dim* sorted subpopulation relaxed into the brighter *Mid* region.

In parallel, analogous sorting experiments were performed with LGIT-infected Jurkats. Cells sorted from the *Off* region of GFP fluorescence remained stably *Off*, whereas cells sorted from the *Bright* region relaxed over time into the *Off* region (Fig. 2d). *Bright*-to-*Off* relaxation kinetics are further explored below. However, this phenomenon appeared distinct from previously observed retroviral silencing (Jaenisch et al., 1982) since LG sorted populations did not exhibit significant relaxation kinetics (Figs. 2b and 3f), and the reduced acetylation K50A Tat mutant exhibited faster relaxation (Fig. 5d), an effect explained below.

In contrast, LGIT cells sorted from the *Mid* region exhibited a novel *Mid*-to-*Bright* relaxation (Fig. 2h-i) where the *Mid* population became significantly brighter over ~10 days, as determined by a Chi-squared test. LGIT-infected cells sorted from the *Dim* region, which corresponds to low Tat levels (see Fig. 3h), produced the most interesting GFP expression dynamics. Seven days after sorting, the LGIT *Dim* sorted cells trifurcated into 3 GFP expression peaks that progressively evolved over time (Fig. 2e). Initially, ~30% of LGIT *Dim* sorted cells were in the *Bright* region, and the remaining 70% were distributed evenly between the *Dim* and *Off* regions. Over a period of 3 weeks, however, the remaining *Dim* subpopulation completely disappeared to leave only the *Bright* and *Off* subpopulations (Fig. 2f). These relaxation dynamics confirmed that the *Dim* region was unstable (Fig. 1b). In contrast, LG *Dim* sorts did not exhibit any *Mid*-to-*Bright* relaxation or *Dim* region instability (Fig. 2b).

LGIT clones sorted from low Tat/GFP concentrations exhibit Phenotypic Bifurcation (PheB)

We have proposed that low basal expression rates may, after random intervals of time, produce small bursts of Tat synthesis subsequently amplified by Tat positive feedback to yield transient and unstable pulses of Tat expression. This hypothesis is supported by the expression dynamics of polyclonal populations of infected cells (Fig. 2). To examine cell populations with uniform basal expression rates, the dynamics of clonal LGIT-infected populations were next investigated (Fig. 1a). Individual cells were sorted from the *Dim* region of an LGIT-infected Jurkat culture, and the 30% that survived sorting were expanded for ~3 weeks. 75% of the resulting 150 clones exhibited GFP fluorescence in the *Off* region, whereas 2% of clones were *Bright* (Fig. 3a). Strikingly, ~23% of clones exhibited a variegated GFP profile. Despite being genetically identical and having a single integration of the LGIT vector at unique chromosomal loci (Fig. 4a), these populations exhibited two subpopulations, one with *Bright* and the other with *Off* fluorescence (Fig. 3b-c). That is, a single LGIT clonal genotype split into two distinct phenotypes, which we term phenotypic bifurcation (PheB). By contrast, clones sorted from the LGIT *Bright* and the LG *Dim* regions produced only single expression phenotypes (Fig. 3d-f). These results are consistent with the interpretation that integration positions supporting zero basal expression stay off, positions supporting relatively high basal expression are rapidly driven on, and clones with intermediate basal rates can fluctuate or exhibit PheB.

Importantly, PheB cells in the *Off* region could be fully transactivated by overnight incubation in TNFα, phorbol myristate acetate (PMA), exogenous recombinant Tat protein, or trichostatin A (TSA) (Fig. 3g). Thus, the HIV LTR in the PheB *Off* subpopulation was intact and still capable of responding to Tat implying that *Off* cells in PheB clones resided in a transcriptionally quiescent but reversible state. In other words, PheB was abrogated either by increasing the Tat concentration through exogenous Tat addition or by increasing the basal rate through TNFα stimulation. Also, in agreement with previous results (Jordan et al., 2001), LG clones responded to, but could not be fully transactivated (into the *Bright* region), by TNFα, PMA, or TSA, whereas Tat incubation could fully transactivate these clones (Supp. Info.).

To demonstrate that GFP expression from LGIT was an accurate reporter of Tat levels, a 2-reporter analog of LGIT was constructed using the monomeric red fluorescent protein (mRFP) (Campbell et al., 2002) and a Tat-GFP fusion protein. Jurkat cells infected with the resulting LTR-mRFP-IRES-TatGFP lentiviral vector demonstrated a linear correlation between mRFP and GFP fluorescence by flow cytometry (Fig. 3h).

Previously, others have utilized an elegant 2-reporter approach to study intrinsic and extrinsic noise contributions in a gene expression system (Elowitz et al., 2002; Raser and O'Shea, 2004). By comparing the relative expression of the two independent reporters, the correlated variability or noise extrinsic to the promoter (i.e. cellular processes, cell division, cell cycle) was distinguished from the uncorrelated noise intrinsic to the specific promoter. Unfortunately, this approach cannot be applied to a closed feedback loop such as Tat transactivation, because each locus would feed back and affect the state of the other, thus destroying the reporter independence the method relies upon. Fortunately, we were otherwise able to directly measure and eliminate the possibility that many noise-generating cellular processes (cell cycle, cell volume, aneuploidy, unequal mitotic division, Tat secretion, growth rate selection, and DNA methylation) contributed to PheB in our LGIT Jurkat system (Supp. Info). Therefore, noise intrinsic to the Tat transactivation loop appeared to be the most parsimonious explanation for PheB, but molecular fluctuations in endogenous transactivation species (e.g. CDk9, Cyclin T1) may also contribute to noise in our system. Thus, our results appear consistent with ideas that eukaryotic noise is gene specific (Raser and O'Shea, 2004) or translation mediated (Ozbudak et al., 2002).

Importantly, LGIT PheB cells that underwent additional *Dim* region sorting rapidly relaxed into the *Off* and *Bright* regions recapitulating a PheB phenotype in the first several days after sorting (Fig. 3i). *Bright* and *Off* sorted LGIT PheB cells were much more stable (Fig. 3i), thereby demonstrating the inherent instability of the *Dim* region and low Tat concentrations.

PheB integration patterns are distinct from non-PheB

Next, the integration sites of clonal populations of LGIT- and LG-infected cells were examined using a nested PCR technique (Schroder et al., 2002). As a byproduct, this analysis also verified that each LGIT PheB clone carried a single, unique lentiviral integration in its genome and confirmed that the low MOI infections yielded single integration events (Fig. 4a). To analyze correlations between integration site and expression dynamics, i.e. genotype and phenotype, a small set of 8 integration sites (4 PheB and 4 non-PheB clones) was initially characterized using the BLAST–like Alignment Tool (BLAT). Multivariate ANOVA testing was conducted on these 8 integration positions to analyze their proximity to several genetic elements previously studied (Schroder et al., 2002): short interspersed nuclear repeats (SINEs), long interspersed nuclear repeats (LINEs), human endogenous retroviral long terminal repeats (HERV LTRs), and genes.

The ANOVA revealed a statistical bias for PheB integrations within 1 kilobase of HERV LTRs ($P \approx 0.05$). To further assess statistical significance, an additional 45 PheB clones were then analyzed for viral integration near HERV LTRs (Supp. Info). This larger data set revealed a higher incidence of HERV-proximate PheB integrations (7/17) that was statistically significantly different from the number of non-PheB clone integrations near HERV LTRs (1/18) ($P \approx 0.02$ by Fisher's Exact 2-tailed Test and $P \approx 0.05$ by Chi-Squared test). Furthermore, the cumulative binomial distribution (Stevens and Griffith, 1994) indicates that the likelihood of integration near HERVs in 7 out of 17 PheB integrations by random chance was extremely low ($P < 10^{-4}$). Integration results are summarized in Fig. 4b.

Notably, PheB clones also had a statistically much higher incidence of integration near HERV LTRs than the much larger (N=524) HIV integration set collected by Schroder et al. (2002) (Fisher's Exact: $P \approx 10^{-4}$ and Chi-Squared: $P < 10^{-7}$). Furthermore, no statistical tests used could differentiate the non-PheB integration set from the much larger Schroder et al. (2002) HIV-1 integration set. This finding suggests that the non-PheB phenotype may be typical of the majority of HIV integration events, and that PheB clones differ significantly from this majority. We also tested whether PheB depended on the proximity of integration to a gene expressed in Jurkat cells and found no correlation (we thank E. Verdin & F. Bushman for Jurkat microarray data, see Supp. Info.).

Heterochromatin has been previously associated with proviral latency (Jordan et al., 2003), and it is possible that HERV LTR sequences variably modulate the local chromatin environment within a clonal population. To assess whether this PEV-like mechanism could drive PheB in some clones, we used exogenous Tat to assay for HIV-1 LTR accessibility. All PheB clones tested could be transactivated by exogenous Tat (Fig. 4c); thus, RNAPII had transcribed TAR, and the HIV-1 LTR was not transcriptionally inaccessible due to heterochromatin modification. By contrast, transcriptionally inactive HIV integrants in dense heterochromatic regions could not be activated by exogenous Tat addition (Jordan et al., 2003). To directly test whether PEV-like mechanisms were affecting PheB integration sites we used quantitative real-time reverse-transcript PCR (qRT-PCR) to measure the transcription level of the genes nearest to three LGIT integrations (clones E7-2, G9-1, and C8-1). The LGIT PheB clones chosen for analysis integrated directly inside, or adjacent to, the corresponding human gene (LAT1-3TM, FoxK2, and C11orf23). No difference in expression of the nearest genes was found between cells isolated from the *Bright* region vs. the *Off* region of these three PheB clones (Fig. 4d).

Furthermore, previous microarray analysis had determined that the expression level of genes in the vicinity of PheB integrations was not altered by the chromatin modifier TSA, indicating that these genes were not susceptible to PEV-like mechanisms (we thank E. Verdin for providing microarray data). Taken together, these data indicate that PEV-like mechanisms do not drive PheB.

Stochastic fluctuations in Tat account for PheB *in silico*

Integration analysis revealed a correlation between integration site and expression phenotype, and we performed computational analysis of the Tat gene expression circuit to examine which specific molecular properties could underlie all cases of PheB. Fig. 4b indicated that PheB integrations were more predisposed to intergenic regions of the genome (Smit, 1999) and may therefore have low basal rates of LTR expression. In fact, the majority of HIV-1 integrations appear to exhibit relatively low basal LTR expression (Jordan et al., 2003) (Supp. Info.), further supporting the hypothesis that Tat exists at low concentrations in PheB clones.

Slow reactions rates that occur among molecules in low copy number make gene expression a highly noisy process. To test if stochastic fluctuations resulting from low basal expression from a single proviral genome coupled with transient amplification by the Tat feedback loop were sufficient to explain PheB, computer models of the transactivation circuit were constructed, and stochastic simulations (Gillespie, 1977) were performed using published parameter values (Reddy and Yin, 1999). Only the simple chemical reactions of Tat transactivation were modeled (Karn 2000; Kaehlcke et al. 2003), and PEV-like mechanisms were not invoked. Alternative model hypotheses (cooperative vs. non-cooperative Tat feedback, Tat acetylation/de-acetylation, high vs. low basal rate, and pre-integration/pre-basal Tat transcription) led to 16 different models that could be compared to experimental data (Table I). Even a qualitative comparison could distinguish the different models from the observations, and only one model class (Eqs. 1-13) successfully generated PheB and matched the experimental data:

$$LTR \xrightarrow{k_{BASAL}} LTR + mRNA_{nuclear} \qquad [1]$$

$$mRNA_{nuclear} \xrightarrow{k_{EXPORT}} mRNA_{cytoplasmic} \qquad [2]$$

$$mRNA_{cytoplasmic} \xrightarrow{k1_{TRANSLATE}} GFP + mRNA_{cytoplasmic} \qquad [3]$$

$$mRNA_{cytoplasmic} \xrightarrow{k2_{TRANSLATE}} Tat_{de-acetylated} + mRNA_{cytoplasmic} \qquad [4]$$

$$Tat_{de-acetylated} \underset{k_{UNBIND}}{\overset{k_{BIND}}{\leftrightarrow}} pTEFb_{de-aceytylated} \qquad [5,6]$$

$$LTR + pTEFb_{de-aceytylated} \underset{k_{DE-ACETYL}}{\overset{k_{ACETYL}}{\leftrightarrow}} pTEFb_{aceytylated} \qquad [7,8]$$

$$pTEFb_{aceytylated} \xrightarrow{k_{TRANSACT}} LTR + mRNA_{nuclear} + Tat_{de-acetylated} \qquad [9]$$

$$GFP \xrightarrow{d_{GPF}} 0,\ Tat_{deacetylated} \xrightarrow{d_{Tat}} 0,\ mRNA_{cytoplasmic} \xrightarrow{d_{CYT}} 0,\ mRNA_{nuclear} \xrightarrow{d_{NUC}} 0 \quad [10\text{-}13]$$

Eqs. 1-13 represent the simplest biologically plausible molecular model that can reproduce PheB. Many molecules known to be important in transactivation (e.g. CDk9, CyclinT1, 7SK, SirT1, p300) were assumed to be constant and "lumped" into effective rate constants. Eqs. 1-13 describe LGIT nuclear mRNA generation from the LTR at a low basal rate, transport to the cytoplasm, and translation into GFP and Tat. Tat then complexes with host factors (assumed to be in stoichiometric excess) to produce pTEFb. pTEFb then binds reversibly to the LTR and is reversibly acetylated (Ott et al., 2004). The acetylated pTEFb stimulates LGIT mRNA production at a high transactivated rate and subsequently releases the bound Tat molecule in the deacetylated form, equivalent to Tat being rapidly deacetylated (Ott et al., 2004). Importantly, the model dictates that Tat degradation is independent of transactivation and that Tat molecules can be recycled through this cycle (Pagans et al., 2005). In addition, simulations were conducted with a distribution of initial Tat concentrations ranging from no Tat to low initial Tat levels, as further described below (Nightingale et al., 2004).

Importantly, Tat/pTEFb acetylation and deacetylation steps (Eqs. 7-8), which functioned as delays in viral transactivation, were required for the simulations to exhibit GFP accumulation kinetics equivalent to those observed after LGIT infection of Jurkat cells (Figs. 2-3), which occurred over 6 days (Table I). In addition, ensuring that $k_{DE\text{-}ACTYL} \gg k_{ACTYL}$ in Eq. 6, supported by *in vitro* data showing that acetylated Tat is rapidly converted to deacetylated Tat after micro-injection into Jurkat cells (Ott et al., 2004), was vital to generating PheB. The resulting acetylation/deacetylation cycle (Eqs. 5-6 & 7-8) created molecular eddies where simulations could pause before either LTR transactivation or Tat/GFP decay.

Low $k_{BASAL}$ values, coupled with an initial or pre-existing Tat concentration of 5-50 molecules, were necessary to generate PheB (Fig. 5a) and accurately reproduce LGIT *Dim* and *Mid* relaxation kinetics (Figs. 2e-i). By contrast, high $k_{BASAL}$ values did not yield PheB and instead produced simulations in which every trajectory very rapidly turned *Bright* (Table I). As previously mentioned, experimental data show that the HIV LTR typically transcribes at a relatively low $k_{BASAL}$ rate after integration (Jordan et al., 2003) (Supp. Info.). Since Tat protein is likely not packaged into virions (Fields et al., 2001), the initial Tat concentration necessary to yield PheB must be produced before the establishment of the low basal transcription rate (i.e. before nuc-1 plants onto the proviral LTR), or even before proviral integration (i.e. pre-integration transcription). Significant evidence for HIV-1 pre-integration transcription exists (Wu, 2004) including the fact that Tat transcripts can be detected within one hour of infection, even in the absence of integration (Stevenson et al., 1990).

Cooperative feedback is known to induce multi-stability (Lai et al., 2004), and despite experimental evidence that argues against Tat cooperativity (Kaehlcke et al., 2003), we tested this possibility by analyzing several models that required two molecules of Tat to fully transactivate the LTR (Table I). In the physiological parameter regime tested, Tat cooperativity produced trajectories that showed continual transitions from *Off* to *Bright*, inconsistent with experimental results (Fig. 2d).

To summarize, PheB could be generated *in silico* by considering only the Tat acetylation cycle, pre-integration transcription, and a low $k_{BASAL}$ value. Although both simpler and more complex versions of Eqs. 1-13 were explored (Table I), these alternate models were incapable of generating PheB under any parameters tested. The results of Table I were not qualitatively affected by perturbing parameters 2-3 logs (i.e. sensitivity analysis) relative to all other parameters (data not shown).

<u>Stochastic model predicts PheB decay and the dynamics of a mutated Tat circuit</u>

Eqs. 1-13 predict that Tat acetylation dynamics play a vital role in determining the average lifetime of cells in the transactivated mode, and this property provided an opportunity to experimentally test the predictive capabilities of the stochastic model. The effect of perturbing the Tat acetylation rate ($k_{ACETYL}$ in Eq. 7) *in silico* was examined, and moderate perturbations in $k_{ACETYL}$ (i.e. ≤ 30%) yielded no change in the *Bright* peak position but led to highly significant changes in the rate of decay from *Bright* to *Off* (Fig. 5b). This prediction was experimentally tested using a Tat mutant containing a lysine-to-alanine (K→A) point mutation at amino-acid 50, a position whose acetylation by p300 is important for transactivation (Kaehlcke et al., 2003). This single K50A point-mutation attenuates acetylation, whereas the related double Tat mutation (K28A+K50A) is lethal to the virus (Kiernan et al., 1999) and thus could not be utilized in this experiment. Infection of Jurkat cells with the LGIT K50A vector generated a *Bright* peak equivalent to "wild-type" LGIT (Fig. 5c); however, sorts collected from the LGIT K50A *Bright* peak decayed significantly more quickly than a parallel LGIT *Bright* sort, in agreement with simulation results (Fig. 5d). Eqs. 1-13 also correctly predicted a second aspect of expression dynamics, the decay rates of different LGIT *Bright* sub-regions (i.e. relative transactivation lifetimes, Supp. Info.).

## DISCUSSION

An integrated experimental and computational study (Fig. 1a) was conducted to explore whether stochastic fluctuations in HIV-1 Tat expression could yield distinct phenotypes analogous to productive and latent HIV-1 infection. Specifically, we explored if proviral integrations into genomic positions that support a low basal expression rate could, after random durations, generate bursts of Tat that were amplified by the positive feedback loop to yield highly variable expression levels. Jurkat cells were infected at low a MOI with an LTR-GFP-IRES-Tat (LGIT) lentiviral vector (Fig. 1c), a model of the Tat transactivation feedback loop. LGIT subpopulations were FACS sorted, and subpopulations that corresponded to low Tat concentrations (i.e. *Dim* GFP fluorescence) were unstable and trifurcated into populations with *Bright*, *Dim*, or *Off* GFP levels (Fig. 2e-f). Clonal populations were generated from the unstable *Dim* region, and 23% of these bifurcated to simultaneously exhibit *Bright* GFP expression and *Off* GFP expression (Fig. 3b-c) despite the fact that all cells in a given clone contained single, unique LGIT proviral integrations in the human Jurkat genome (Fig. 4a). Noise in cellular processes extrinsic to this Tat transactivation circuit, including cell cycle, cell volume, aneuploidy, Tat secretion, and variegated local gene expression did not account for this phenotypic bifurcation (Fig. 4d, and Supp. Info.). However, the genomic integration sites of PheB clones statistically differed from those of non-PheB clones (Fig. 4b) and may have generated low basal transcription rates. Lastly, *in silico* computational models of Tat transactivation suggested that Tat stochastics generated PheB (Fig. 5 and Table I), and the model correctly predicted the experimentally observed decay dynamics of a K50A Tat mutant.

PheB demonstrates that latent and transactivated lentiviral modes are alternate probabilistic expression modes, and stochastic molecular fluctuations driving such a variegated phenotype have not been previously reported in lentiviruses or mammalian cells. PheB appears to be the outcome of a molecular decision between the completion or disruption of the Tat transactivation circuit (Fig. 6a). In this model, a small pre-integration transcriptional burst produces a few Tat molecules that enter a cycle of probabilistic pTEFb acetylation and deacetylation. After a random interval of time, if pTEFb can "run the molecular gauntlet" and escape this rapid cycle to transactivate the LTR, Tat is produced, the circuit is completed, and viral gene expression is strongly activated. However, it should be emphasized that this genetic circuit is classically monostable. The high-expressing, transactivated mode is merely a long-lived pulse resulting from transient positive feedback amplification of small fluctuations in Tat expression, and the stronger back-reactions (de-acetylation, degradation and unbinding) eventually deactivate the circuit. Thus, this long-lived stochastic Tat transactivation pulse, given enough time, eventually decays in this model system. Importantly, once Tat transactivation occurs with wild type HIV-1, viral production rapidly ensues and kills infected cells within 1.7 days (Perelson et al., 1996).

The model predicted that perturbing the forward or reverse reaction rates would alter transactivation dynamics. Indeed, the K50A Tat acetylation mutant, which would experience decreased forward reaction rates, displayed decreased transactivation lifetimes (Fig. 5d). In further support of this model, experiments that increased the forward reaction rates or transactivation efficiency by downregulating the endogenous pTEFb inhibitor HEXIM1 (Yik et al., 2004) abrogated PheB, increased fluorescence of the Bright subpopulation, and, most significantly, increased transactivation lifetime (Supp. Info).

Another key aspect of the model in Fig. 6a is the inclusion of the phenomenon of pre-integration transcription. Retroviral pre-integration transcription is not widely cited, for a review see (Wu, 2004), but Tat transcripts are detectable within one hour of infection, even in the absence of integration (Stevenson et al., 1990). Transcription without integration has also been detected in human primary CD4 infected with wild-type HIV-1 (Wu, 2004) and in integrase-defective lentiviral vector infections in culture (Nightingale et al., 2004). Pre-integration transcription implies that the time-to-proviral-integration may be one stochastic process leading to PheB and therefore potentially latency, since the amount of time the HIV-1 pre-integration complex remains un-integrated could influence the amount of pre-transcribed Tat. Importantly, pre-integration transcription is only necessary to explain PheB, a case when the integration dependent basal rate is predicted to be exceptionally low. Integrations near Alu elements or genes endowed with a relatively high basal rate would rapidly induce a strongly-transactivated circuit and overshadow the effect of a small, early transcriptional burst.

There are several areas for further potential development of this work. Eqs. 1-13 present a minimal model that is capable of predicting and gaining molecular insights into the process of phenotypic bifurcation, and analogous modeling approaches have been extremely successful in describing complex biological phenomena, such as HIV-1 *in vivo* dynamics (Perelson et al., 1996). Future efforts may include greater molecular detail and additional comparison of models to data. Second, although a continuously growing body of fundamental work on HIV propagation, gene expression dynamics, and regulation has been conducted with Jurkat cells (Pagans et al., 2005), they are a leukemic T cell line. However, we also have evidence of PheB-like behavior in other cells lines (SupT1 and HeLa cells). These observations could be explored in primary CD4 T cell cultures, but it is uncertain whether they will exhibit PheB-like HIV-1 expression since primary T cell cultures can only be maintained in culture in a transcriptionally over-activated state (via anti-CD3 or phytohaemagglutinin) that would stimulate high basal HIV-1 expression rates. In support of this assertion, 293 kidney epithelial cells, known for high levels of NF-κB activation (Horie et al., 1998) and thus high HIV-1 LTR activation, did not exhibit relaxation kinetics or PheB (Supp. Info). It should also be noted that while Tat stochastics could play a role in delaying HIV-1 transactivation and contributing to latency, reactivation of HIV-1 from latency, analogous to TNFα stimulation, would strongly upregulate HIV-1 gene expression and abrogate PheB.

Future studies are also needed to explore the correlation between PheB and integration near HERV LTRs. HERVs occupy ~1% of the human genome and are believed to be the remnants of ancient germ-cell retroviral infections that occurred 10-60 million years ago (Sverdlov, 2000). HERVs have been reported to exert epigenetic influence over gene expression via DNA methylation (Okada et al., 2002). Although DNA methylation does not appear to affect the HIV-1 LTR (Jordan et al., 2001; Pion et al., 2003), the effect of histone methylation, known to be dependent upon DNA methylation (He and Lehming, 2003), on the HIV-1 LTR has not yet been reported, and this chromatin modification may influence transcription. Nevertheless, the findings that all PheB clones could be transactivated by exogenous Tat protein (Fig 4c), that expression of genes near LGIT integrations was not affected by TSA, and that transcription from these nearby genes was equivalent in *Bright* and *Off* PheB populations (Fig. 4d) argues against chromatin modification and PEV-like mechanisms acting in this system.

HIV has been shown to preferentially integrate into transcriptionally active regions including Alu repeats (Schroder et al., 2002). Although highly speculative, we propose it may be in the best evolutionary interest of HIV-1 to integrate into regions that support productive (non-PheB) expression, but integrate with low frequency into the remaining intergenic regions of the genome that permit only a low basal expression and may induce PheB. The latter integrations could be fully capable of viral production, but long stochastic delays before production could be interrupted by conversion of a T cell to a memory state in vivo to generate a latent virus (Fig. 6b). In this way HIV-1 could 'hedge its bets' and have a robust mechanism to survive aggressive host immune responses. Bacteriophage λ uses a similar probabilistic mechanism to regulate lysis-lysogeny status during periods of poor host nutrition and thus for an opportune nutritional window to lyse its host and ensure optimal progeny propagation.

## METHODS

### Cloning & viral production

DNA manipulations were performed using standard techniques, and GFP refers to enhanced GFP (Clontech, Palo Alto, CA). Briefly, LTR-GFP (LG) was constructed by deleting the CMV promoter from the plasmid CMV-LTR-CMV-GFP (Miyoshi et al., 1998). The 2 exon version of *tat* utilized in the LTR-GFP-IRES-Tat (LGIT) was obtained from pEV680 (a kind gift from Eric Verdin, UCSF). The final GIT cassette within the LGIT vector was sequenced for verification. LTR-mRFP-IRES-TatGFP (LRITG) was generated from LGIT by exchanging *gfp* for the monomeric Red Fluorescent Protein (*mRFP*) (Campbell et al., 2002), a kind gift from Roger Tsien, and by swapping *tat* for a *tat-gfp* fusion (Qbiogene Inc., Carlsbad CA). The K50A Tat mutant was a kind gift from Melanie Ott and was inserted in place of Tat in GIT. Cloning details are available upon request.

Lentiviral vectors were packaged and concentrated as previously described (Dull et al., 1998), yielding between $10^7$ and $10^8$ infectious units/ml, as determined by flow cytometry analysis of infected Jurkat cells after TNFα incubation.

Cell Culture & Cytometry

Jurkat cells were maintained in RPMI 1640 with L-glutamine (Invitrogen, Carlsbad, CA), 10% fetal bovine serum, and 1% penicillin-streptomycin, at concentrations between $10^5$ and $2\times10^6$ cells/mL under humidity and 10% $CO_2$ at 37°C. Jurkat cells were infected by incubation with concentrated lentiviral vector overnight in the presence of 8 μg/ml polybrene. After 2 days, cultures were analyzed by flow cytometry on a Beckman-Coulter EPICS XL-MCL cytometer (http://biology.berkeley.edu/crl/cell_sorters_analysers.html). All flow measurements were performed in parallel with an uninfected Jurkat control and statistically analyzed using FlowJo (Treestar Inc., Ashland Oregon). Infected cultures were stimulated in parallel after 4 days by PMA, TNF-α, TSA, or exogenous Tat protein as previously described (Jordan et al., 2001) to ensure that <8% of cells were GFP+ (MOI~0.08) (Fields et al., 2001).

7-12 days after infection, GFP+ cells were sorted on a Beckman-Coulter EPICS Elite ESP Sorter. Both bulk population, polyclonal sorts and single cell, monoclonal sorts were conducted for a range of different GFP fluorescence regions, including the "*Dim*" (~1-10 RFU), "*Mid*" (~10-50 RFU), "*Bright*" (~50-1024 RFU), and "*Off*" regions (between 0.1-1 RFU).

See Supp. Info. for microscopy sample preparation.

Integration Site Analysis and qRT-PCR

Integration sites were determined and analyzed as previously described (Schroder et al., 2002) (see Supp. Info for primer sequences and detailed methods). A clonal integration site was accepted only if it satisfied 3 criteria: (1) a single band >500 bp observed on a 1.5% agarose gel, (2) chromatographic sequencing results showing a single sequence, and (3) homology to the human genome (BLAT hit) beginning at the HIV-1 LTR/human genome junction and having sequence identity of >95% for at least 90 nucleotides. The UCSC genome browser (http://genome.ucsc.edu/, 07/03 assembly) was then used to identify sequences elements surrounding the integration sites. See Supp. Info for detailed statistical procedures and qRT-PCR details.

Computer Modeling

Stochastic simulations were performed by the direct simulation method of the Chemical Master Equation (Gillespie, 1976), and C++ code was adapted from (Lai et al., 2004). GFP trajectories and histograms were plotted with PLPLOT (http://www.plplot.org). EGFP calibration beads (Clontech, Palo Alto, CA) were used to determine the molecules-to-RFU conversion: EGFP molecules = 37700*RFU – 4460 for our Beckmann Coulter cytometer.

## ACKNOWLEDGEMENTS

We thank Eric Verdin and Melanie Ott, for generously donating reagents and for many helpful discussions, Jasper Rine & Michael Botchan for helpful discussions and critical review of this manuscript, Rob Rodick (DPE Group) for generous donation of computer time and extensive computer support, and Qiang Zhou for the 7SK plasmid. We are indebted to Pat Flaherty for statistical expertise and thank Josh Leonard for reagents and Nory Cabanilla for experimental assistance. We also thank M. Roberts, S. Shroff, T. Altman, & S. Andrews for general technical assistance. FACS was performed at the UC Berkeley Flow Cytometry Core Facility, and we thank Hector Nolla for extensive technical support. APA was supported by the Howard Hughes Medical Institute and the Defense Advanced Research Projects Agency. DVS was supported by an NSF CAREER Award. LSW was supported by a pre-doctoral fellowship from the Howard Hughes Medical Institute.

## REFERENCES

Arkin, A., Ross, J., and McAdams, H. H. (1998). Stochastic kinetic analysis of developmental pathway bifurcation in phage lambda-infected Escherichia coli cells. Genetics *149*, 1633-1648.

Becskei, A., Seraphin, B., and Serrano, L. (2001). Positive feedback in eukaryotic gene networks: cell differentiation by graded to binary response conversion. Embo J *20*, 2528-2535.

Becskei, A., and Serrano, L. (2000). Engineering stability in gene networks by autoregulation. Nature *405*, 590-593.

Blake, W. J., M, K. A., Cantor, C. R., and Collins, J. J. (2003). Noise in eukaryotic gene expression. Nature *422*, 633-637.

Brooks, D. G., Kitchen, S. G., Kitchen, C. M., Scripture-Adams, D. D., and Zack, J. A. (2001). Generation of HIV latency during thymopoiesis. Nat Med *7*, 459-464.

Campbell, R. E., Tour, O., Palmer, A. E., Steinbach, P. A., Baird, G. S., Zacharias, D. A., and Tsien, R. Y. (2002). A monomeric red fluorescent protein. Proc Natl Acad Sci U S A *99*, 7877-7882.

Dull, T., Zufferey, R., Kelly, M., Mandel, R. J., Nguyen, M., Trono, D., and Naldini, L. (1998). A third-generation lentivirus vector with a conditional packaging system. J Virol *72*, 8463-8471.

Elowitz, M. B., and Leibler, S. (2000). A synthetic oscillatory network of transcriptional regulators. Nature *403*, 335-338.

Elowitz, M. B., Levine, A. J., Siggia, E. D., and Swain, P. S. (2002). Stochastic gene expression in a single cell. Science *297*, 1183-1186.

Fields, B. N., Knipe, D. M., and Howley, P. M. (2001). Fields' virology, 4th edn (Philadelphia, Lippincott Williams & Wilkins).

Finzi, D., Blankson, J., Siliciano, J. D., Margolick, J. B., Chadwick, K., Pierson, T., Smith, K., Lisziewicz, J., Lori, F., Flexner, C., *et al.* (1999). Latent infection of CD4+ T cells provides a mechanism for lifelong persistence of HIV-1, even in patients on effective combination therapy. Nat Med *5*, 512-517.

Gillespie, D. T. (1976). General Method for Numerically Simulating Stochastic Time Evolution of Coupled Chemical-Reactions. Journal of Computational Physics *22*, 403-434.

Gillespie, D. T. (1977). Exact Stochastic Simulation of Coupled Chemical-Reactions. Journal of Physical Chemistry *81*, 2340-2361.

Hanahan, D., Lane, D., Lipsich, L., Wigler, M., and Botchan, M. (1980). Characteristics of an SV40-plasmid recombinant and its movement into and out of the genome of a murine cell. Cell *21*, 127-139.

He, H., and Lehming, N. (2003). Global effects of histone modifications. Brief Funct Genomic Proteomic *2*, 234-243.

Horie, R., Aizawa, S., Nagai, M., Ito, K., Higashihara, M., Ishida, T., Inoue, J., and Watanabe, T. (1998). A novel domain in the CD30 cytoplasmic tail mediates NFkappaB activation. Int Immunol *10*, 203-210.

Isaacs, F. J., Hasty, J., Cantor, C. R., and Collins, J. J. (2003). Prediction and measurement of an autoregulatory genetic module. Proc Natl Acad Sci U S A *100*, 7714-7719.

Jaenisch, R., Harbers, K., Jahner, D., Stewart, C., and Stuhlmann, H. (1982). DNA methylation, retroviruses, and embryogenesis. J Cell Biochem *20*, 331-336.

Jordan, A., Bisgrove, D., and Verdin, E. (2003). HIV reproducibly establishes a latent infection after acute infection of T cells in vitro. Embo J *22*, 1868-1877.

Jordan, A., Defechereux, P., and Verdin, E. (2001). The site of HIV-1 integration in the human genome determines basal transcriptional activity and response to Tat transactivation. Embo J *20*, 1726-1738.

Kaehlcke, K., Dorr, A., Hetzer-Egger, C., Kiermer, V., Henklein, P., Schnoelzer, M., Loret, E., Cole, P. A., Verdin, E., and Ott, M. (2003). Acetylation of Tat defines a cyclinT1-independent step in HIV transactivation. Mol Cell *12*, 167-176.

Karn, J. (2000). Tat, a novel regulator of HIV transcription and latency. In HIV Sequence Compendium 2000, C. Kuiken, F. McCutchan, B. Foley, M. JW, H. B, M. J, M. P, and S. M. Wolinsky, eds. (Los Alamos, NM, Theoretical Biology and biophysics Group, Los Alamos National Laboratory), pp. 2-18.

Kiernan, R. E., Vanhulle, C., Schiltz, L., Adam, E., Xiao, H., Maudoux, F., Calomme, C., Burny, A., Nakatani, Y., Jeang, K. T.*, et al*. (1999). HIV-1 tat transcriptional activity is regulated by acetylation. Embo J *18*, 6106-6118.

Kutsch, O., Benveniste, E. N., Shaw, G. M., and Levy, D. N. (2002). Direct and quantitative single-cell analysis of human immunodeficiency virus type 1 reactivation from latency. J Virol *76*, 8776-8786.

Lai, K., Robertson, M. J., and Schaffer, D. V. (2004). The sonic hedgehog signaling system as a bistable genetic switch. Biophys J *86*, 2748-2757.

Lin, X., Irwin, D., Kanazawa, S., Huang, L., Romeo, J., Yen, T. S., and Peterlin, B. M. (2003). Transcriptional profiles of latent human immunodeficiency virus in infected individuals: effects of Tat on the host and reservoir. J Virol *77*, 8227-8236.

McAdams, H. H., and Arkin, A. (1997). Stochastic mechanisms in gene expression. Proc Natl Acad Sci U S A *94*, 814-819.

Miyoshi, H., Blomer, U., Takahashi, M., Gage, F. H., and Verma, I. M. (1998). Development of a self-inactivating lentivirus vector. J Virol *72*, 8150-8157.

Mizuguchi, H., Xu, Z., Ishii-Watabe, A., Uchida, E., and Hayakawa, T. (2000). IRES-dependent second gene expression is significantly lower than cap-dependent first gene expression in a bicistronic vector. Mol Ther *1*, 376-382.

Nightingale, S., Hollis, R. P., Yang, C., Bahner, I., Pepper, K. A., and Kohn, D. B. (2004). Transient Gene Expression by Non-Integrating Lentiviral (NIL) Vectors. Molecular Therapy *9*, S159.

Okada, M., Ogasawara, H., Kaneko, H., Hishikawa, T., Sekigawa, I., Hashimoto, H., Maruyama, N., Kaneko, Y., and Yamamoto, N. (2002). Role of DNA methylation in transcription of human endogenous retrovirus in the pathogenesis of systemic lupus erythematosus. J Rheumatol *29*, 1678-1682.

Ott, M., Dorr, A., Hetzer-Egger, C., Kaehlcke, K., Schnolzer, M., Henklein, P., Cole, P., Zhou, M. M., and Verdin, E. (2004). Tat acetylation: a regulatory switch between early and late phases in HIV transcription elongation. Novartis Found Symp *259*, 182-193; discussion 193-186, 223-185.

Ozbudak, E. M., Thattai, M., Kurtser, I., Grossman, A. D., and van Oudenaarden, A. (2002). Regulation of noise in the expression of a single gene. Nat Genet *31*, 69-73.

Pagans, S., Pedal, A., North, B. J., Kaehlcke, K., Marshall, B. L., Dorr, A., Hetzer-Egger, C., Henklein, P., Frye, R., McBurney, M. W., *et al.* (2005). SIRT1 Regulates HIV Transcription via Tat Deacetylation. PLoS Biol *3*, e41.

Perelson, A. S., Neumann, A. U., Markowitz, M., Leonard, J. M., and Ho, D. D. (1996). HIV-1 dynamics in vivo: virion clearance rate, infected cell life-span, and viral generation time. Science *271*, 1582-1586.

Pierson, T., McArthur, J., and Siliciano, R. F. (2000). Reservoirs for HIV-1: mechanisms for viral persistence in the presence of antiviral immune responses and antiretroviral therapy. Annu Rev Immunol *18*, 665-708.

Pion, M., Jordan, A., Biancotto, A., Dequiedt, F., Gondois-Rey, F., Rondeau, S., Vigne, R., Hejnar, J., Verdin, E., and Hirsch, I. (2003). Transcriptional suppression of in vitro-integrated human immunodeficiency virus type 1 does not correlate with proviral DNA methylation. J Virol *77*, 4025-4032.

Raser, J. M., and O'Shea, E. K. (2004). Control of stochasticity in eukaryotic gene expression. Science *304*, 1811-1814.

Reddehase, M. J., Podlech, J., and Grzimek, N. K. (2002). Mouse models of cytomegalovirus latency: overview. J Clin Virol *25 Suppl 2*, S23-36.

Reddy, B., and Yin, J. (1999). Quantitative intracellular kinetics of HIV type 1. AIDS Res Hum Retroviruses *15*, 273-283.

Reuter, G., and Spierer, P. (1992). Position effect variegation and chromatin proteins. Bioessays *14*, 605-612.

Schroder, A. R., Shinn, P., Chen, H., Berry, C., Ecker, J. R., and Bushman, F. (2002). HIV-1 integration in the human genome favors active genes and local hotspots. Cell *110*, 521-529.

Smit, A. F. (1999). Interspersed repeats and other mementos of transposable elements in mammalian genomes. Curr Opin Genet Dev *9*, 657-663.

Spudich, J. L., and Koshland, D. E., Jr. (1976). Non-genetic individuality: chance in the single cell. Nature *262*, 467-471.

Stevens, S. W., and Griffith, J. D. (1994). Human immunodeficiency virus type 1 may preferentially integrate into chromatin occupied by L1Hs repetitive elements. Proc Natl Acad Sci U S A *91*, 5557-5561.

Stevenson, M., Stanwick, T. L., Dempsey, M. P., and Lamonica, C. A. (1990). HIV-1 replication is controlled at the level of T cell activation and proviral integration. Embo J *9*, 1551-1560.

Sverdlov, E. D. (2000). Retroviruses and primate evolution. Bioessays *22*, 161-171.

Wu, Y. (2004). HIV-1 gene expression: lessons from provirus and non-integrated DNA. Retrovirology *1*, 13.

Yik, J. H., Chen, R., Pezda, A. C., Samford, C. S., and Zhou, Q. (2004). A human immunodeficiency virus type 1 Tat-like arginine-rich RNA-binding domain is essential for HEXIM1 to inhibit RNA polymerase II transcription through 7SK snRNA-mediated inactivation of P-TEFb. Mol Cell Biol *24*, 5094-5105.

**Figure Legends**

Figure 1: Schematic vectors and flowchart. (a) A flowchart of the logical and experimental progression followed in this study of the importance of stochastic fluctuations in Tat transactivation. (b) Heuristic summary of results from a preliminary stochastic model (Supp. Info) showing the bi-modality of the transactivation circuit as a function of Tat concentration for a provirus with a low basal expression rate. *Off* and *Bright* (transactivated) expression modes exist, with the *Bright* mode appearing less stable. At low Tat concentrations, a highly unstable shoulder or knife-edge caused trajectories to diverge into either *Bright* (transactivated) or *Off* modes. (c) The HIV genome is compared to the transactivation-model vector LTR-GFP-IRES-Tat (LGIT). In both HIV and LGIT, Tat can transactivate the LTR, increasing transcriptional elongation and completing a strong positive feedback loop. However, at low Tat concentrations stochastic fluctuations can influence completion of the feedback loop.

Figure 2: GFP expression histograms of *in vitro* and *in silico* FACS sorted subpopulations of LTR-GFP-IRES-Tat (LGIT) infected Jurkats. (a) LTR-GFP (LG) infected Jurkats 7 days after infection (4% are GFP+). (b) LG *Dim* and *Mid* bulk sorts ($10^4$ cells sorted) were analyzed 1 month post-sorting, and no bifurcation or relaxation kinetics were observed. (c) LGIT-infected Jurkats 7 days after infection (6.5% are GFP+). (d) LGIT *Off* and *Bright* bulk sorts 7 days post-sorting. Post-sort analysis confirmed 98% sorting fidelity (data not shown). (e) GFP expression dynamics of LGIT *Dim* bulk sorted cells (black outline) together with a computer-simulated *Dim* sort (solid grey) of Eqs. 1-13 (see text). *Dim* sorted cells trifurcated in GFP expression after 7 days, such that ~30% of cells remained *Dim*, ~30% switched *Off* and, most strikingly, ~30% turned *Bright*, as seen more easily in the time-course (f). 35 days later the remaining *Dim*

population had completely relaxed into only *Off* and *Bright* populations. Stochastic simulations (initiated with ≈ 50,000 GFP molecules) successfully reproduced the trifurcation and relaxation dynamics. (g) *Mid* sorts relaxed into the *Bright* region over 20 days, as seen in the 3D overlay of histograms. Notably, the GFP-axis is measured on a log-scale, and there is a significant difference in GFP between days 7 and 20 (by a Chi-squared test), as seen more easily in the time-course (h). (i) Conversion of the LGIT *Mid* sort histograms into a color-map representation (the histogram peak is depicted as pink, and each row is a different day) shown together with a grey-scale color-map of an *in silico* LGIT *Mid* bulk sort (using Eqs. 1-13 initiated with ≈ 300,000 GFP molecules).

Figure 3: Clonal populations generated from LTR-GFP-IRES-Tat (LGIT) and LTR-GFP (LG) infections of Jurkat cells. (a) Proportions of phenotypes exhibited by clonal populations generated from FACS sorting of single cells from the LGIT *Dim* GFP region. Of the 30% of *Dim* cells that successfully expanded, 73% generated clones with no/*Off* GFP expression, 2% produced *Bright* clones, and 25% of clones exhibited phenotypic bifurcation (PheB). (b) A representative fluorescence micrograph of an LGIT PheB clone (green = GFP, blue = DAPI nuclear staining). (c) Flow histograms of LGIT PheB clones. (d) Clones sorted from the LGIT *Bright* region do not exhibit PheB. (e) A representative LGIT *Bright* clone micrograph. (f) Clones sorted from the LG *Dim* region do not exhibit PheB. (g) LGIT PheB clones (red) can be fully transactivated by chemical perturbation, including a 17 hour incubation in TNFα (green), TSA (yellow), and exogenous Tat protein (1μg, lightest blue; 6μg, darker blue; and 12.5μg, darkest blue). (h) Infection of cells with the two-reporter control LTR-mRFP-IRES-TatGFP shows a strong correlation between the 1st and 2nd cistrons. (i) Additional FACS sub-sorting from the *Bright*, *Dim*, and *Off* regions of three LGIT PheB clones. For all clones, *Dim* sorted cells rapidly relaxed into the *Off* and *Bright* regions recapitulating a bifurcating phenotype in the first several days after sorting, whereas *Bright* and *Off* sorted cells appeared significantly more stable.

Figure 4: PCR verification of monoclonality and sequencing of LGIT integration sites. (a) Gel electrophoresis of GenomeWalker PCR results for 8 different PheB clones (lanes 2-9), together with PCR controls generated from a polyclonal bulk cell mixture (bulk LGIT *Dim* sort, lane 10) and two clonal populations deliberately mixed together (lane 11). (b) Results of BLAT analysis of integration sites in the human genome. Non-PheB clones integrated into genes ~72% of the time. PheB clones integrated within 1 kb of a HERV LTR ~47% of the time ($P<0.01$). (c) Exogenous Tat alone can significantly transactivate both PheB clones with HERV-proximate integration (clones 1-4) and PheB clones integrated near genes (clones 5-8). (d) Quantitative real-time RT-PCR analysis of gene expression from three LGIT PheB clonal integration sites (clones C8-1, E7-2, and G9-1). For each clone *Bright* and *Off* subpopulations were sorted and transcription of the HIV-1 $\Psi$ region was compared to transcription of the nearest human gene (FoxK2, C11orf23, and LAT1-3TM; C8-1 and E7-2 integrated within the respective genes). Data are reported as the molar ratio (normalized to beta-actin) of *Bright* sort expression to *Off* sort expression for each locus (i.e. fold difference in locus expression between *Bright* and *Off* sort). HIV-1 $\Psi$ expression was always significantly higher in the *Bright* sort as compared to the *Off* sort (*Bright* sort $\Psi$ expression was greater than Off sort $\Psi$ expression by ~100-fold for E7-2, ~10-fold for C8-1 and ~1000-fold for G9-1) but the difference in expression of the nearest human gene was not statistically different in these subpopulations ($P>0.05$).

Table I: Stochastic simulation results for different Tat transactivation model classes. Eqs. 1-13, and 15 variations of these mechanisms, were simulated. Eqs. 1-13 results are shown at the intersection of "*Acetylation Included*" and "*No Tat Cooperativity*", while all other quadrants are model variations that failed to produce PheB. Acetylation refers to the presence or absence of Eqs. 7-8, and Tat Cooperativity models had a modified Eq. 7 that required 2 Tat molecules for transactivation. The latter class of models always resulted in trajectories that eventually turned *Bright* (i.e. *Off* unstable), inconsistent with data (Figs. 2-3). In all simulations, initial GFP concentrations corresponded to a *Dim* sort where $GFP_0 \approx 30,000$ molecules; initial $Tat_0 \geq 1$ indicates pre-integration transcription; and crossed-out quadrants are logically unhelpful for mechanism discrimination.

Figure 5: *In silico* modeling of PheB and model-based prediction of mutant (Tat K50A) LGIT dynamics. (a) Eqs. 1-13 were used to simulate a trajectory of GFP over 3 weeks of virtual time. This trajectory is "clonal" (the simulation was conducted with a single, constant set of parameter values, Supp. Info). This clonal stochastic simulation resulted in a PheB histogram, shown in the inset. (b) Model-based predictions of the dynamics of a hypothetical LGIT vector containing a Tat mutant. An LGIT circuit defective in acetylation (~30% lower forward-acetylation rate, Eq. 7) exhibited an identical *Bright* peak position but accelerated decay from *Bright* to *Off*. (c) Experimental verification of model-based prediction: FACS histogram of Jurkats cells infected with LGIT (grey) and the LGIT K50A (black) reduced acetylation mutant. The *Bright* peak position is equivalent. (d) Experimental confirmation of model-based prediction: relaxation dynamics of *Bright* sorts from Jurkat cells infected with LGIT (grey data points) are slower than those of cells infected with LGIT K50A (black data points). Upper and lower 95% confidence

intervals (dashed lines of corresponding color) are shown, and corresponding simulations, from (b), are shown as solid lines. Experimentally measured slopes of the LGIT and K50A decay rates were statistically different ($P<0.0001$ by regression analysis with interaction variable) and this result was confirmed by statistical jack-knifing of the data.

Figure 6: Molecular model for PheB and its potential relationship to HIV-1 proviral latency (a) A molecular model of the Tat transactivation circuit. Basal LTR expression is low, and a small pre-integration transcriptional burst produces a few molecules of Tat that "jump start" the transactivation circuit. Tat/pTEFb molecules then run a "molecular stochastic gauntlet" where back-reactions in the circuit form "eddies" that cause Tat to pause at intermediate concentrations. pTEFb can exit the eddies either by decaying or by initiating positive feedback and transactivating the LTR. (b) A stochastic decision may contribute to latency. *In vivo*, HIV-1 integrations supporting a high basal rate will rapidly transactivate. However, a low basal rate could behave like PheB: HIV-1 could transactivate after a random time delay, or memory cell conversion could generate a latent virus first.

Fig. 1

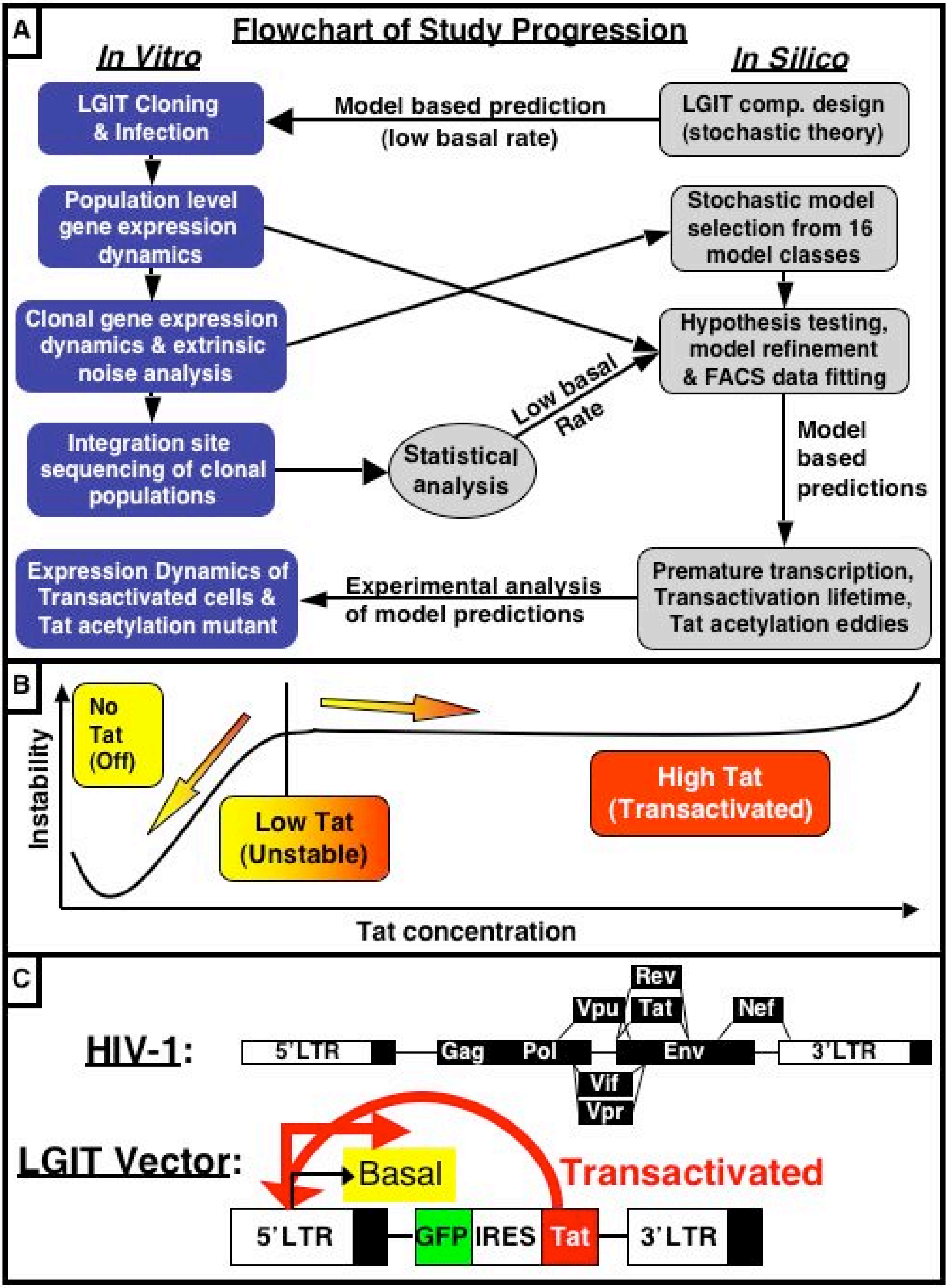
A
Flowchart of Study Progression
In Vitro
In Silico
LGIT Cloning & Infection
Model based prediction (low basal rate)
LGIT comp. design (stochastic theory)
Population level gene expression dynamics
Stochastic model selection from 16 model classes
Clonal gene expression dynamics & extrinsic noise analysis
Hypothesis testing, model refinement & FACS data fitting
Low basal Rate
Integration site sequencing of clonal populations
Statistical analysis
Model based predictions
Expression Dynamics of Transactivated cells & Tat acetylation mutant
Experimental analysis of model predictions
Premature transcription, Transactivation lifetime, Tat acetylation eddies
B
No Tat (Off)
Low Tat (Unstable)
High Tat (Transactivated)
Instability
Tat concentration
C
HIV-1:
5'LTR
Gag
Pol
Vpu
Rev
Tat
Nef
Env
Vif
Vpr
3'LTR
LGIT Vector:
Basal
Transactivated
5'LTR
GFP
IRES
Tat
3'LTR

# Fig. 2

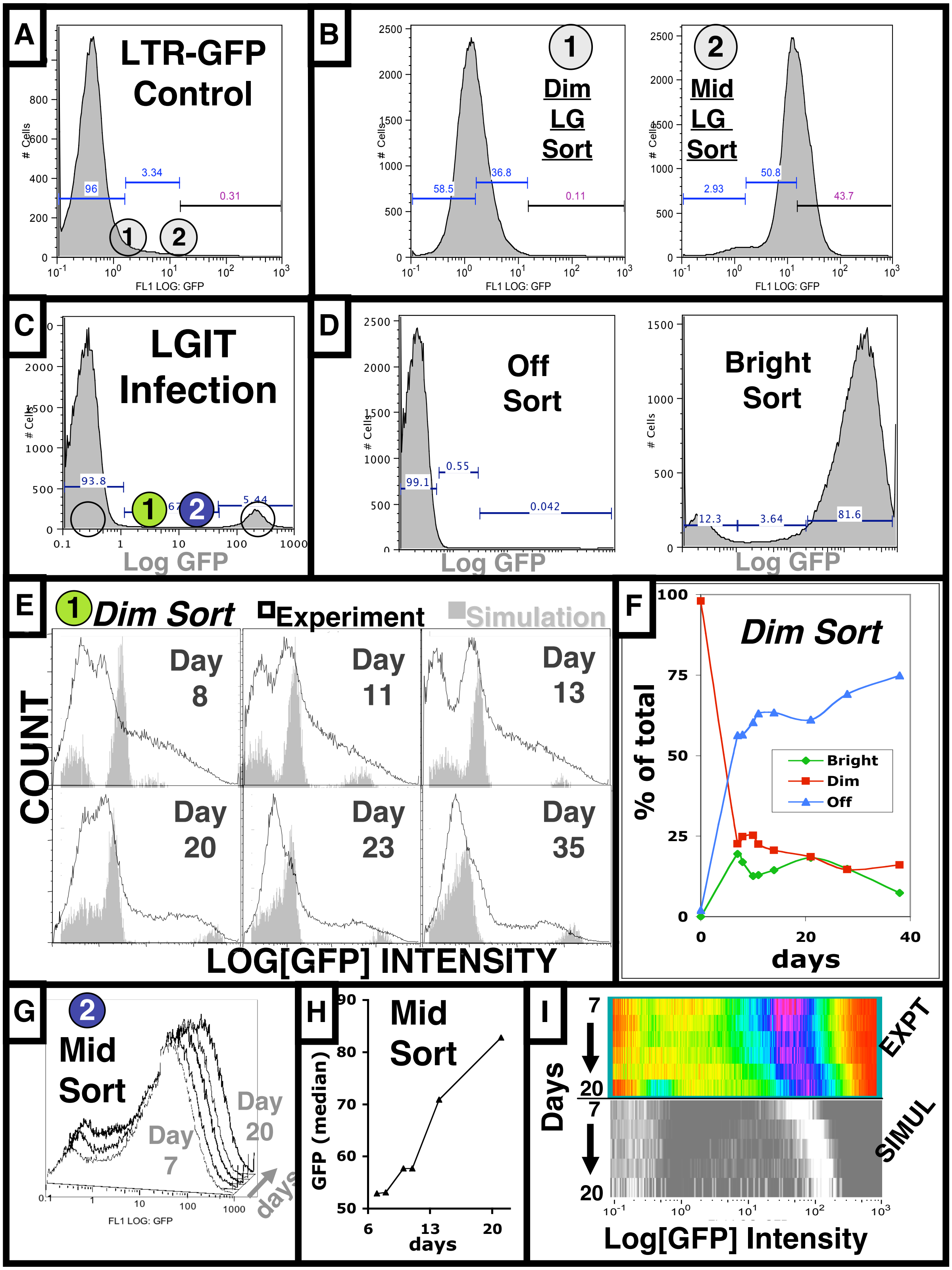

# Fig. 3

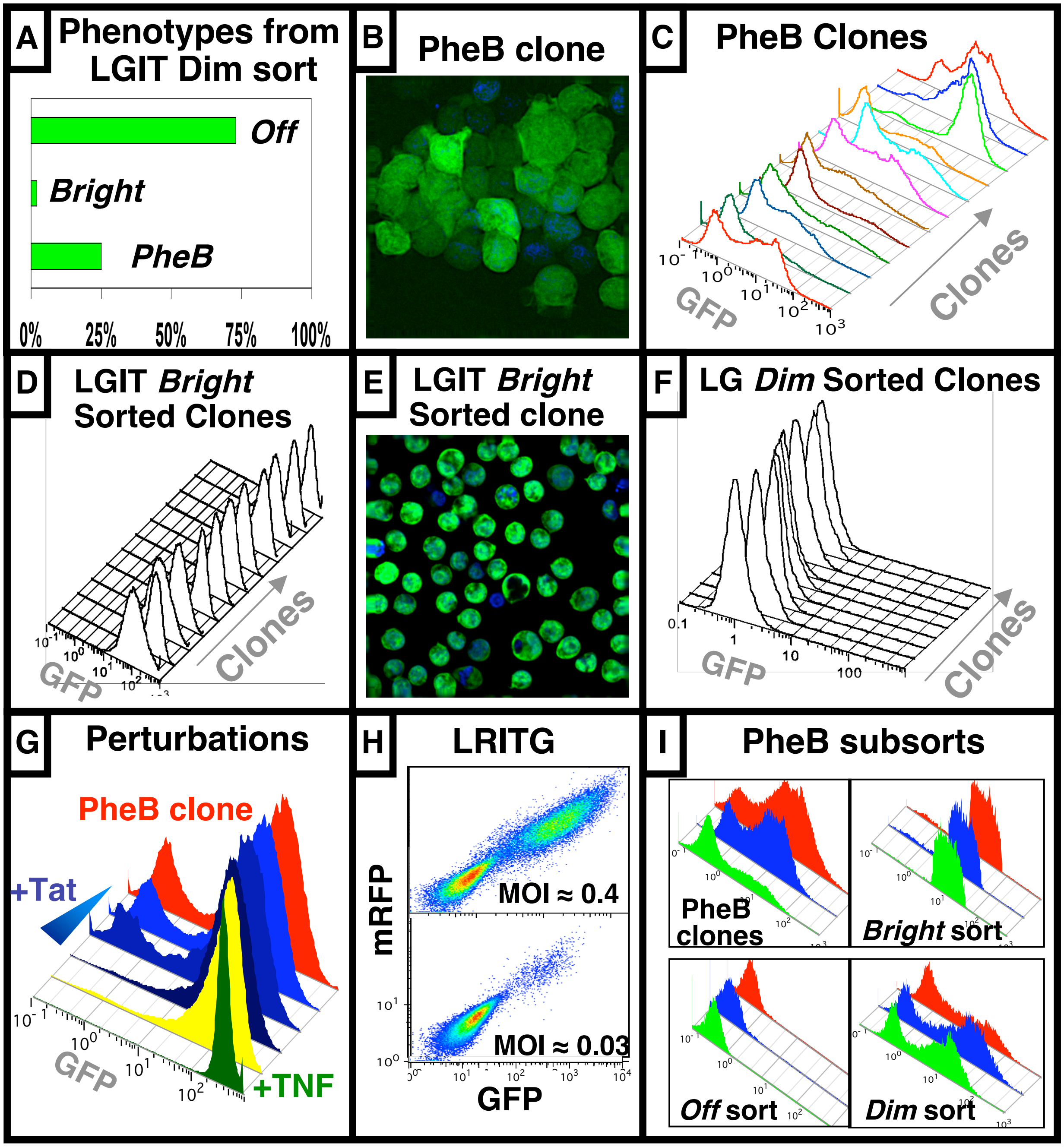

Fig. 4

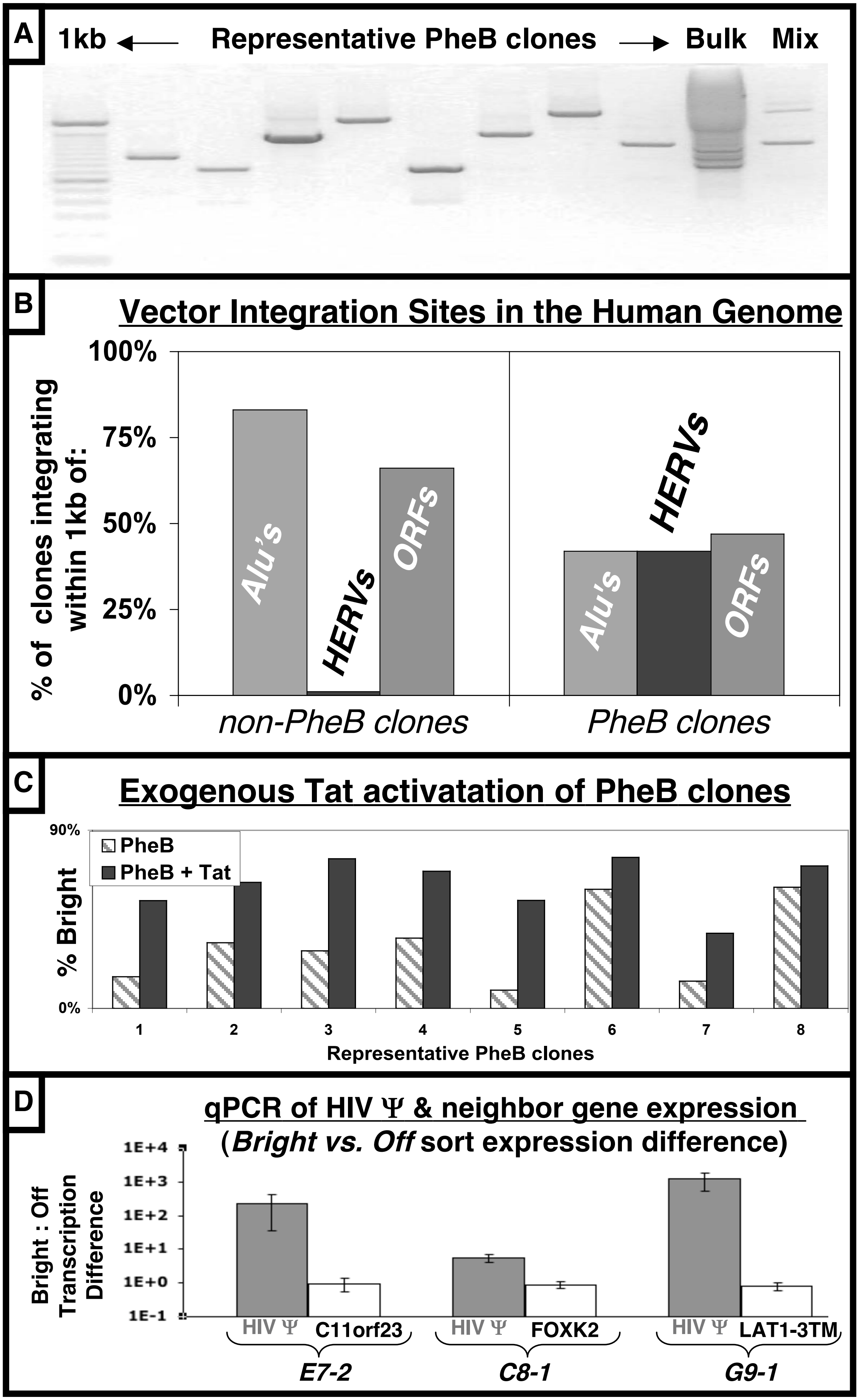
A
1kb
Representative PheB clones
Bulk
Mix
B
Vector Integration Sites in the Human Genome
% of clones integrating within 1kb of:
100%
75%
50%
25%
0%
Alu's
HERVs
ORFs
non-PheB clones
PheB clones
C
Exogenous Tat activatation of PheB clones
90%
0%
PheB
PheB + Tat
% Bright
1
2
3
4
5
6
7
8
Representative PheB clones
D
qPCR of HIV Ψ & neighbor gene expression
(Bright vs. Off sort expression difference)
Bright : Off Transcription Difference
1E+4
1E+3
1E+2
1E+1
1E+0
1E-1
HIV Ψ
C11orf23
FOXK2
LAT1-3TM
E7-2
C8-1
G9-1

**Table I**:

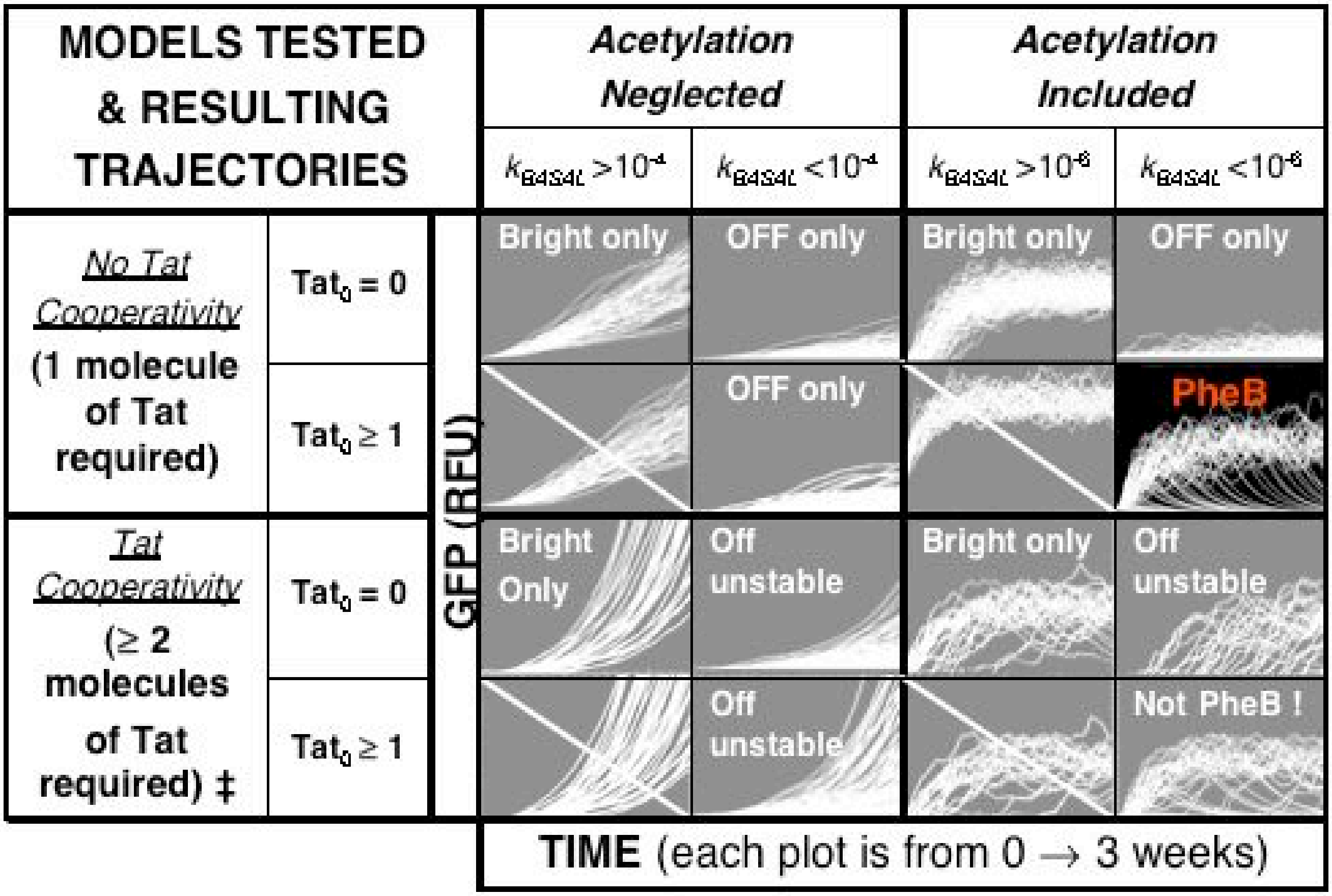
MODELS TESTED & RESULTING TRAJECTORIES
Acetylation Neglected
Acetylation Included
$k_{BASAL}$ >10^-4
$k_{BASAL}$ <10^-4
$k_{BASAL}$ >10^-6
$k_{BASAL}$ <10^-6
No Tat Cooperativity
(1 molecule of Tat required)
Tat0 = 0
Tat0 ≥ 1
Tat Cooperativity
(≥ 2 molecules of Tat required) ‡
GFP (RFU)
Bright only
OFF only
Bright only
OFF only
OFF only
PheB
Bright Only
Off unstable
Bright only
Off unstable
Off unstable
Not PheB !
TIME (each plot is from 0 → 3 weeks)

# Fig. 5

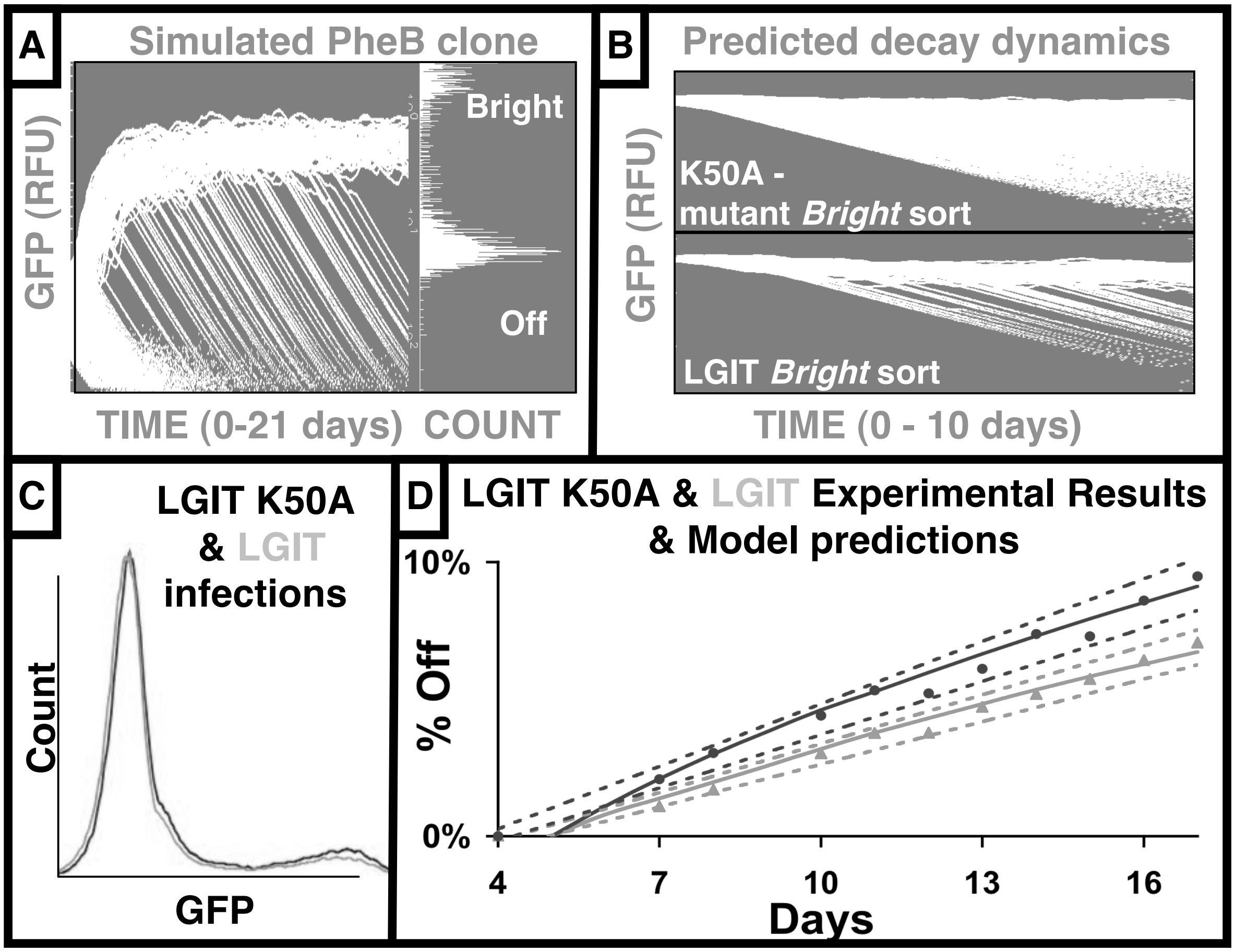

Fig. 6

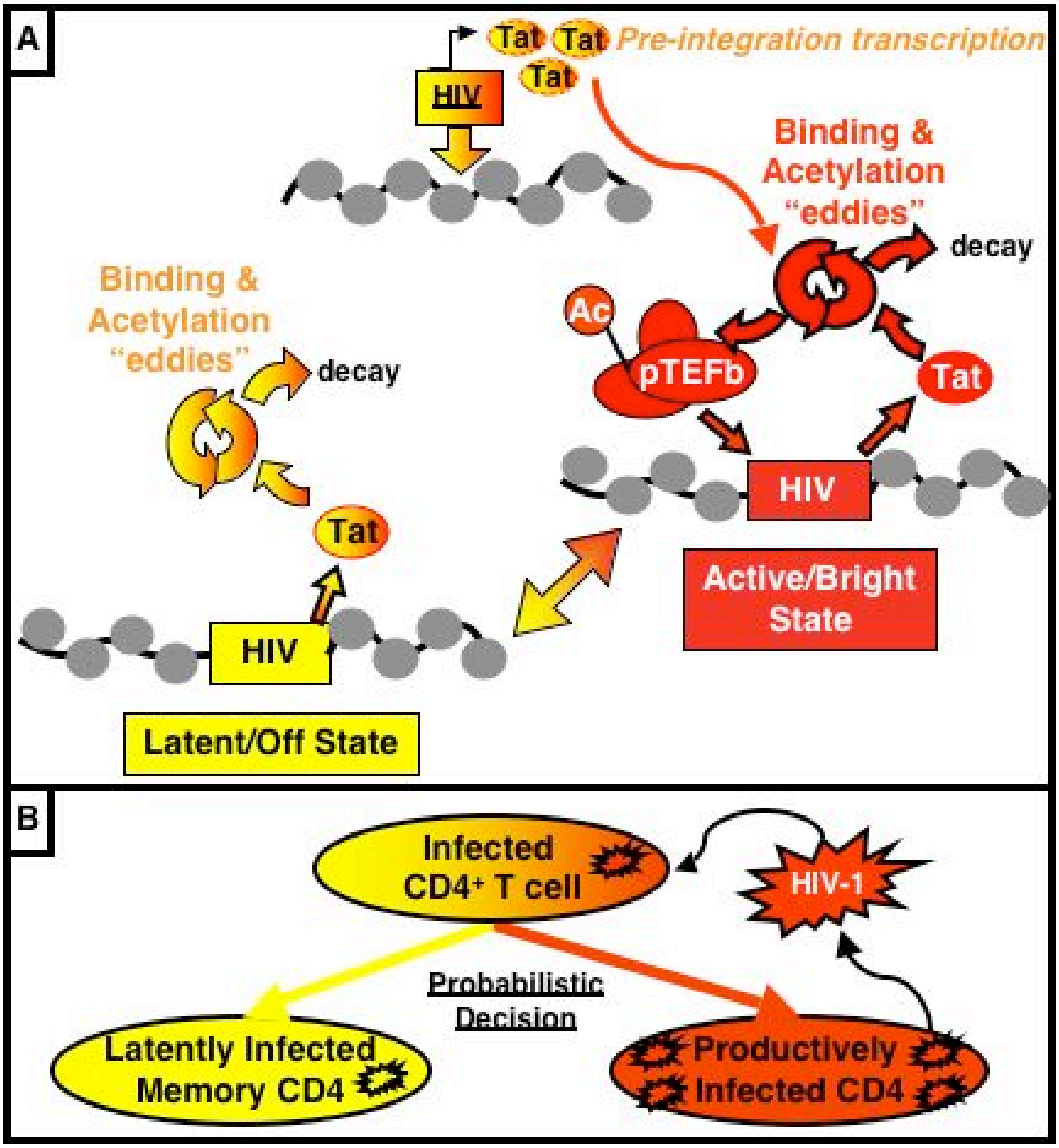